\documentclass[twocolumn, nofootinbib, 
prd, floats, floatfix, amsmath, amssymb, 
superscriptaddress, preprintnumbers]{revtex4-2} %
\UseRawInputEncoding
\usepackage[final]{graphicx}
\usepackage{hyperref}
\usepackage{amsmath}
\usepackage{bbm}
\usepackage{bm}
\usepackage{amsfonts}
\usepackage{amssymb}
\usepackage{latexsym}
\usepackage{graphicx}
\usepackage[english]{babel}
\usepackage{multirow}
\usepackage{float}
\usepackage{url}
\usepackage{slashed}
\usepackage{xcolor} 
\usepackage[utf8]{inputenc}
\usepackage{stmaryrd} 
\usepackage{enumitem}
\usepackage{hyperref}
\usepackage{cleveref}
\usepackage{siunitx}
\usepackage{verbatim}
\usepackage{orcidlink}
\usepackage{amsthm}
\usepackage{booktabs}
\usepackage{graphicx}
\usepackage{multirow}
\theoremstyle{remark}

\newcommand{\be}{\begin{equation}}
\newcommand{\ee}{\end{equation}}
\newcommand{\ba}{\begin{array}}
\newcommand{\ea}{\end{array}}
\newcommand{\bea}{\begin{eqnarray}}
\newcommand{\eea}{\end{eqnarray}}

\newcommand{\besub}{\begin{subequations}}
\newcommand{\eesub}{\end{subequations}}

\newcommand{\beq}{\begin{equation} \begin{aligned}}
		\newcommand{\eeq}{\end{aligned} \end{equation}}

\definecolor{darkerblue}{rgb}{0.2,0.2,0.5}
\definecolor{seagreen}{rgb}{0.180392,0.545098,0.341176}
\definecolor{smagenta}{rgb}{0.5,0.145098,0.341176}
\definecolor{deepblue}{rgb}{0,0,1}

\begin{document}

\title{Topological defects as effective dynamical dark energy}
	
	\author{Haipeng An}
	\email{anhp@mail.tsinghua.edu.cn}
	\affiliation{Department of Physics, Tsinghua University, Beijing 100084, China}
	\affiliation{Center for High Energy Physics, Tsinghua University, Beijing 100084, China}
	
	\author{Chengcheng Han \orcidlink{0000-0003-1063-2282}}
	\email{hanchch@mail.sysu.edu.cn}
	\affiliation{School of Physics, Sun Yat-Sen University, Guangzhou 510275, P. R. China}
    \affiliation{Asia Pacific Center for Theoretical Physics, Pohang 37673, Korea}
	\author{Borui Zhang}
	\email{zhangbr22@mails.tsinghua.edu.cn}
	\affiliation{Department of Physics, Tsinghua University, Beijing 100084, China}

\begin{abstract}
In this work, we consider the possibility that the dynamical dark energy hinted at by recent DESI data may be mimicked by the effects of additional components in the universe, potentially arising from topological defects. We find that the data does not show a particular preference for the existence of cosmic strings. However, a domain wall contribution at the percent level can improve the fit, yielding a $\Delta \chi^2= -1.72$ compared to the $\Lambda \rm{CDM}$ model. The improvement indicates that topological defects remain a viable and interesting extension to $\Lambda$CDM, meriting further investigation with future cosmological data.
\end{abstract}
\maketitle

\section{Introduction}

\begin{figure}[t]
 \centering
   \includegraphics[width=0.4\textwidth]{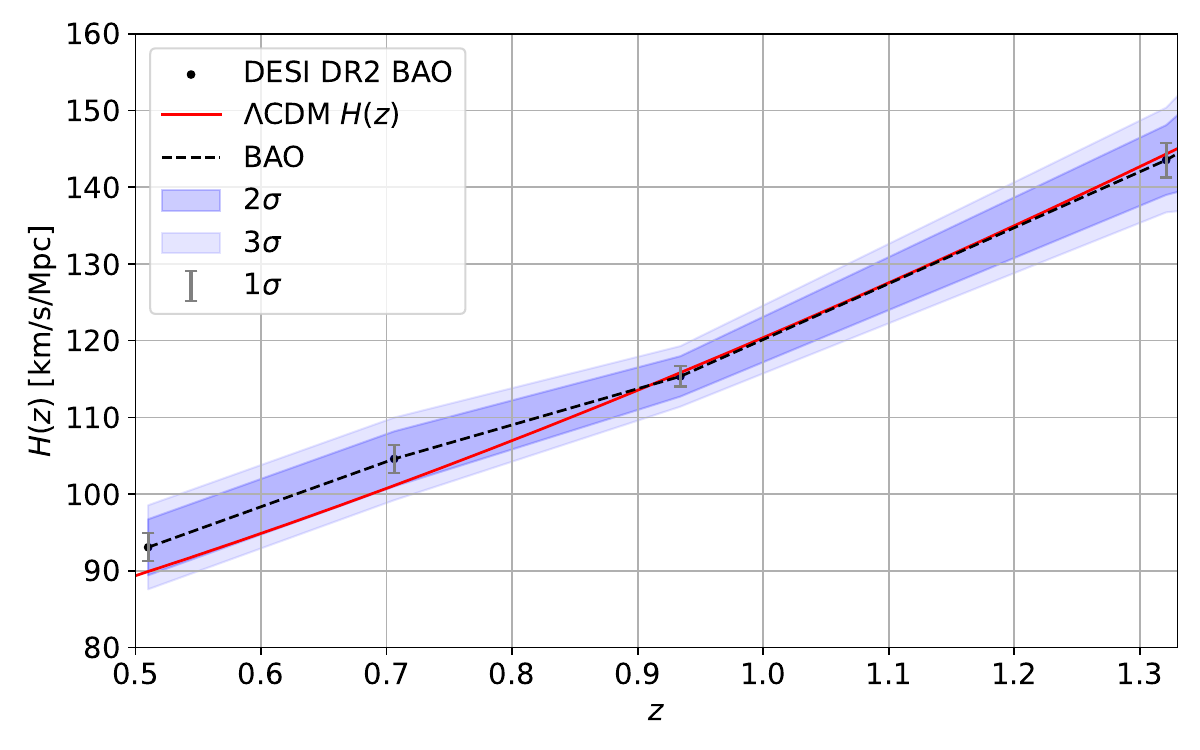}
\caption{Hubble expansion rate measured by the DESI-2 from $z=0.5$ to $z=1.35$.}
\label{fig:Hubble}
\end{figure}

The driver of the accelerating expansion of our Universe is the most mystery in fundamental physics. In the standard model of cosmology, it is provided by the cosmological constant, which is equivalent to the vacuum energy of quantum fields and therefore is also known as the dark energy (DE). In this case, the equation of state parameter for dark energy, $w_{\rm DE} = -1$. The equation of state parameter, $w_{\rm DE}$ can be extracted from the measurement of the evolution of the baryon acoustic oscillation (BAO) using large scale structure surveys. Recent result from Dark Energy Spectroscopic Instrument (DESI) found significant deviation of the Hubble expansion rate from what is predicted by the standard $\Lambda$CDM model in their second data release (DR2)~\cite{DESI:2025zgx,DESI:2025fii,DESI:2025wyn}. As shown in Fig.~\ref{fig:Hubble}, we can see that the measured Hubble expansion rate is larger than what is predicted by the $\Lambda$CDM model (shown by the red curve) at the redshift $z\sim 0.5 - 0.8$. If this deviation is physical, it indicates that energy density in addition to the $\Lambda$CDM model is injected into our Universe in this redshift region. This deviation can be explained by introducing evolving DE with intriguing dynamics~\cite{Guo:2004fq,Hu:2004kh,Huang:2025som,Oriti:2025lwx,Ormondroyd:2025iaf,Luu:2025fgw,Lee:2025yvn,Nakagawa:2025ejs,Anchordoqui:2025fgz,Hur:2025lqc,Pang:2025lvh,Nesseris:2025lke,Chaussidon:2025npr,You:2025uon,Kessler:2025kju,Akrami:2025zlb,Murai:2025msx,deSouza:2025rhv,Dinda:2025iaq,Teixeira:2025czm,Wolf:2025jed,Li:2025cxn,Lin:2025gne,Smith:2025grk,Wang:2025rll,Csillag:2025gnz,Bayat:2025xfr,Moffat:2025jmx,Cai:2025mas}, non-cold dark matter~\cite{Kumar:2025etf,Abedin:2025dis}, dynamical dark matter~\cite{Wang:2025zri,Chen:2025wwn}, dark matter-DE interactions~\cite{Huey:2004qv,Das:2005yj,Aboubrahim:2024cyk,Chakraborty:2025syu,Khoury:2025txd,Shah:2025ayl,Silva:2025hxw,You:2025uon,Pan:2025qwy,Shlivko:2025fgv,Yang:2025mws,Cruickshank:2025iig,vanderWesthuizen:2025iam}, non-vanishing cosmic curvature contribution~\cite{Dinda:2023kvg}, modified theories of gravity from various UV considerations~\cite{Zhu:2024qcm,Pan:2025psn,Paliathanasis:2025dcr,Yang:2025mws,Plaza:2025gjf,Paliathanasis:2025hjw,Scherer:2025esj,Smirnov:2025yru,Postolak:2025qmv}, assuming existence of non-trivial local structures~\cite{Banik:2025dlo,Moffat:2025sik}, modifying the recombination history of the Universe~\cite{Mirpoorian:2025rfp}.

The data suggests an evolving dark energy model with an phantom-like equation of state (EOS) transitioning from $w<-1$ to $w>-1$ in recent times. In this work, instead of invoking an evolving dark energy component, we propose that there may be a missing part in the composition of our universe. In particular, the inclusion of topological defects can mimic the evolving dark energy.

Topological defects, such as cosmic strings and domain walls, evolve differently from dark energy and dark matter. 
For the frustrated domain walls or cosmic strings, the equations of state satisfy,
\begin{align}
	w_{\rm td}=\left\{ \begin{matrix}& -\frac{2}{3},\quad \text{domain wall} \\\\
	&-\frac{1}{3},\quad \text{cosmic string} \end{matrix}\right..
\end{align}

To fit the observed data, we consider the possibility that a non-negligible fraction of the present-day energy density of the Universe is attributed to topological defects. A substantial population of domain walls or cosmic strings is strongly constrained by current measurements of cosmic microwave background (CMB) anisotropies. These constraints typically require the cosmic string tension to satisfy $G\mu < \mathcal O(10^{-7})$~\cite{Planck:2013mgr}, and the phase transition associated with domain wall formation to occur at $\eta< O(1)$ MeV~\cite{Lazanu:2015fua, Sousa:2015cqa}. These conditions translate into upper bounds on the fractional energy densities $\Omega_{\rm cs} < \mathcal O(10^{-7})$ for cosmic strings and $\Omega_{\rm dw} < \mathcal{O}(10^{-4})$ for domain walls.

However, such limits are generally derived under the assumption that the topological defects remain in the scaling regime, where their energy density redshifts efficiently with cosmic expansion. In contrast, frustrated domain walls or cosmic strings redshift more slowly — their energy density dilutes less rapidly than in the scaling case. For a fixed present-day fraction, frustrated defects would have contributed less to the total energy density at earlier times, resulting in weaker constraints from the CMB.
Moreover, frustrated topological defects do not cluster or form large-scale structures, which suppresses their impact on CMB anisotropies. The constraints can be further relaxed if the defects enter the horizon only at late times in cosmic history. This requires the defects to have a sufficiently large correlation length. Taking these effects into account, it remains plausible that topological defects could contribute up to a percent-level fraction of the current energy density of the Universe. In this work, we investigate this possibility and quantify its implications.

\begin{figure*}[t]
	\centering
	\includegraphics[width=0.49\textwidth]{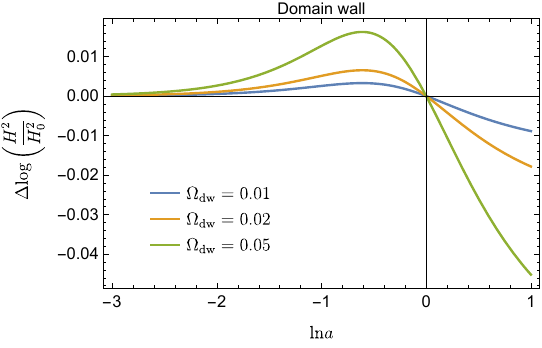}
	\includegraphics[width=0.49\textwidth]{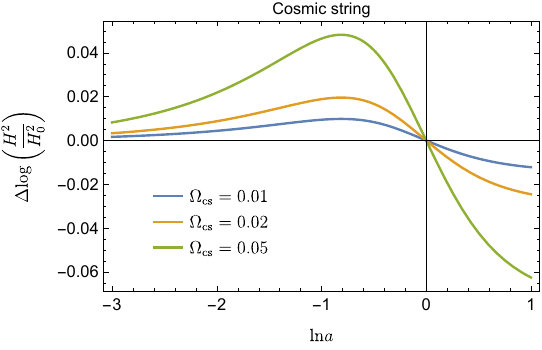}
	\caption{The figure illustrating the difference of $\mathrm{log}\left(H^2/H_0^2\right)$ between topological defects and $\Lambda$CDM.}
	\label{fig:hubble_td}
 \end{figure*}

\section{The evolution of $H(z)$}

In our setup, we assume a flat universe with $\Omega_k=0$. Then the expansion rate of the universe can be written as
\begin{align}
\frac{H^2(z)}{H_0^2}&=\Omega_{\rm bc}(1+z)^3+\Omega_{\gamma}(1+z)^4+\Omega_{\nu}\frac{\rho_\nu(z)}{\rho_{\nu0}}\notag\\
&+\Omega_{\rm td}(1+z)^{3(1+w_{\rm td})}+\Omega_{\Lambda},
\end{align}
where $H_0$ is the Hubble parameter today, $\Omega_{\rm bc} = \Omega_b + \Omega_c$ denotes the current fractional density of baryons and cold dark matter. $\Omega_\gamma$ and $\Omega_\nu$ represent the current fractional densities of radiation and neutrinos, respectively. The total matter density is defined as $\Omega_{\rm m0} = \Omega_{\rm bc} + \Omega_\nu$ when neutrinos are non-relativistic. We adopt the dark energy as the cosmological constant with fraction $\Omega_\Lambda=1-\Omega_{\rm m0}-\Omega_{\rm td}-\Omega_\gamma$.

We plot the ratio of the Hubble parameter of the topological-defect models to that of the \(\Lambda\)CDM model, with \(H_0\) and \(\omega_{b(c)} \equiv \Omega_{b(c)}h^2\) held fixed, as shown in Fig.~\ref{fig:hubble_td}. From the plot, a bump appears whose peak is near the scale factor \(a \simeq 0.55\text{--}0.7\), corresponding to redshift \(z \simeq 0.4\text{--}0.8\), consistent with the Hubble parameter increase measured by DESI DR2, as shown in Fig.~\ref{fig:Hubble}. In our model, the sole free parameter is the present-day fractional density of topological defects. By varying this parameter, we find that the amplitude of the bump increases with increasing defect fraction. This behavior is expected: topological defects redshift faster than the cosmological constant but slower than matter, so they are negligible at early times. However, since their present-day fraction is fixed, their fractional contribution exceeds that of the cosmological constant across a wide range of redshifts, thereby mitigating the decline of the Hubble parameter compared to a model with only matter and a cosmological constant.

To sample posteriors in parameter space, we employ the Metropolis-Hastings Markov Chain Monte Carlo (MCMC) algorithm, using the cosmological inference framework \textbf{Cobaya}~\cite{Torrado:2020dgo} which implements adaptive fast-slow-optimized MCMC samplers for Bayesian analysis. The theoretical predictions are based on a modified version of \textbf{CAMB}~\cite{Lewis:1999bs}, which is integrated within \textbf{Cobaya}. To constrain the model parameters, we utilize measurements from both baryon acoustic oscillations (BAO) and Type Ia supernovae (SNe). BAO data are taken from DESI DR2, while SNe measurements are drawn from three recent datasets: Pantheon+~\cite{Scolnic:2021amr,Brout:2022vxf}, Union3~\cite{Rubin:2023ovl}, and DESY5~\cite{DES:2024jxu}. All these datasets are incorporated into \textbf{Cobaya} as likelihoods.

\begin{table}[h!]
\centering
\begin{tabular}{llc}
\toprule
\textbf{model} & \textbf{parameter} & \textbf{prior} \\
\midrule
\textbf{Baseline} & $\omega_{\mathrm{c}}$ & $\mathcal{U}[0.11, 0.13]$ \\
                  & $\omega_{\mathrm{b}}$   & $\mathcal{U}[0.021, 0.024]$ \\
                  & $100\theta_{\mathrm{MC}}$ & $\mathcal{U}[1.03, 1.05]$ \\
\midrule
	\multicolumn{3}{l}{\textbf{Topological defects}+$\mathbf{\Lambda}$\textbf{CDM}} \\
\textbf{Domain wall} & $\omega_{\rm dw}$ & $\mathcal{U}[0, 0.05]$ \\
\textbf{Cosmic string} & $\omega_{\rm cs}$ & $\mathcal{U}[0, 0.05]$ \\
\midrule
\textbf{$\Omega_k$+$\mathbf{\Lambda}$\textbf{CDM}} & $\Omega_{\rm k}$ & $\mathcal{U}[0, 0.05]$ \\
\midrule
\end{tabular}
\caption{Parameter priors used in various dark energy parametrizations.}
\end{table}

We adopt the CMB-inferred values of $(\theta_*, \omega_b, \omega_{\rm bc})_{\rm CMB}$ as a correlated Gaussian distribution~\cite{DESI:2025zgx}. Correlated Gaussian posteriors on the parameters $(\theta_*, \omega_b, \omega_{\rm bc})$, obtained by marginalizing over the ISW and lensing contributions and other possible late-time effects, are used in our analysis. We incorporate this information in the form of a correlated Gaussian distribution, as a more model-independent and computationally efficient alternative to the full CMB likelihood, which we refer to in the following as $(\theta_*, \omega_b, \omega_{\rm bc})_{\rm CMB}$. We use the result performed by \cite{Lemos:2023xhs,DESI:2025zgx} based on the \textbf{CamSpec} CMB likelihood, which defines a compressed likelihood which can be achieved by the Gaussian mixture likelihood sector implemented in \textbf{Cobaya} with mean
\begin{equation}
\bm{\mu}(\theta_*, \omega_b, \omega_{bc}) =
\begin{pmatrix}
0.01041 \\
0.02223 \\
0.14208
\end{pmatrix},
\end{equation}

and the covariance matrix yields
\begin{equation}\label{cv_CMB}
\mathbf{C} = 10^{-9} \times
\begin{pmatrix}
0.006621 & 0.12444 & -1.1929 \\
0.12444 & 21.344 & -94.001 \\
-1.1929 & -94.001 & 1488.4
\end{pmatrix}.
\end{equation}

Also, We adopt the Gelman–Rubin convergence diagnostic \( R - 1 < 0.01 \)~\cite{Gelman:1992zz} as the stopping criterion for all MCMC chains. After convergence is achieved, the first 30\% of the samples are discarded as burn-in.
 The fitting result is shown in Tab.\,\ref{table1}

Since the measurements of the parameters \( \theta_*, \omega_b \) and \(  \omega_{c} \) are highly precise~\cite{Planck:2018vyg} (in the sampling, $\omega_{bc}$ acts as a derived quantity), and the MCMC sampling results (in terms of \( \chi^2 \)) are not sensitive to the choice of priors, we adopt narrower prior ranges for these three parameters compared to those used in~\cite{DESI:2025fii}. This choice improves sampling efficiency while still ensuring that the prior ranges are sufficiently broad to encompass the Planck measurement uncertainties~\cite{Planck:2018vyg}—specifically, the prior range exceeds the central value by more than \( \pm 5\sigma \)—and the variance of the likelihood in Eq.~\ref{cv_CMB}.

\begin{figure*}[t]
		\centering
		\includegraphics[width=0.5\textwidth]{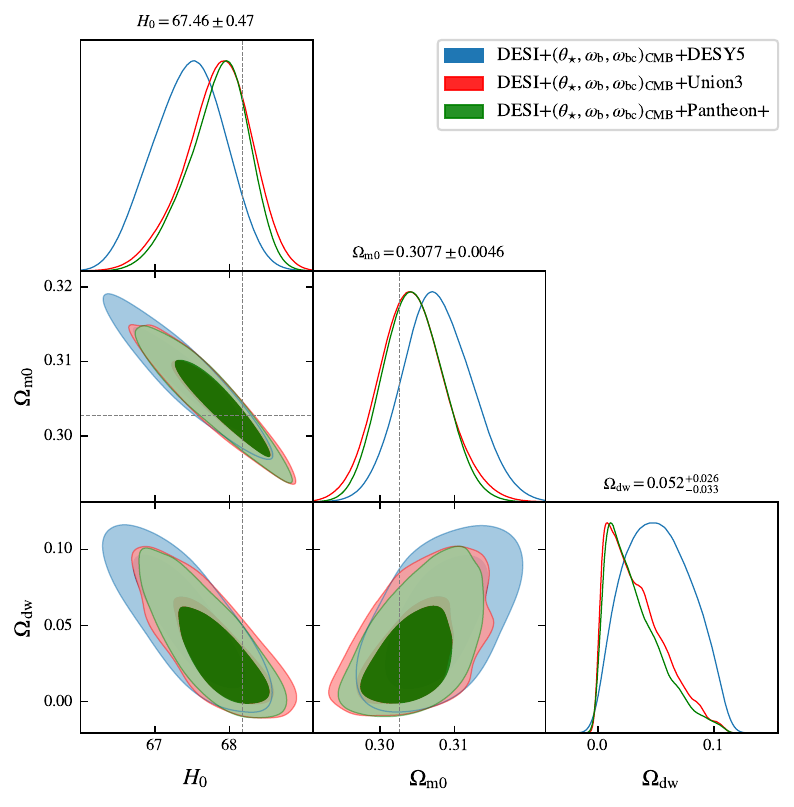}
		\caption{Posterior distributions of parameters in the domain wall model combined with DESI, CMB and supernovae. The model is with one sampling parameter $\Omega_{\rm dw}$. The blue contours indicate the constraint from DESY5, the red contours correspond to Union3 and the green contours are for the Pantheon+. The gray dashed lines correspond to the value of $H_0$ and $\Omega_{\rm m0}$ of $\Lambda$CDM \cite{DESI:2025zgx}.}\label{dw}
\end{figure*}

\begin{figure*}[t]
		\centering
		\includegraphics[width=0.5\textwidth]{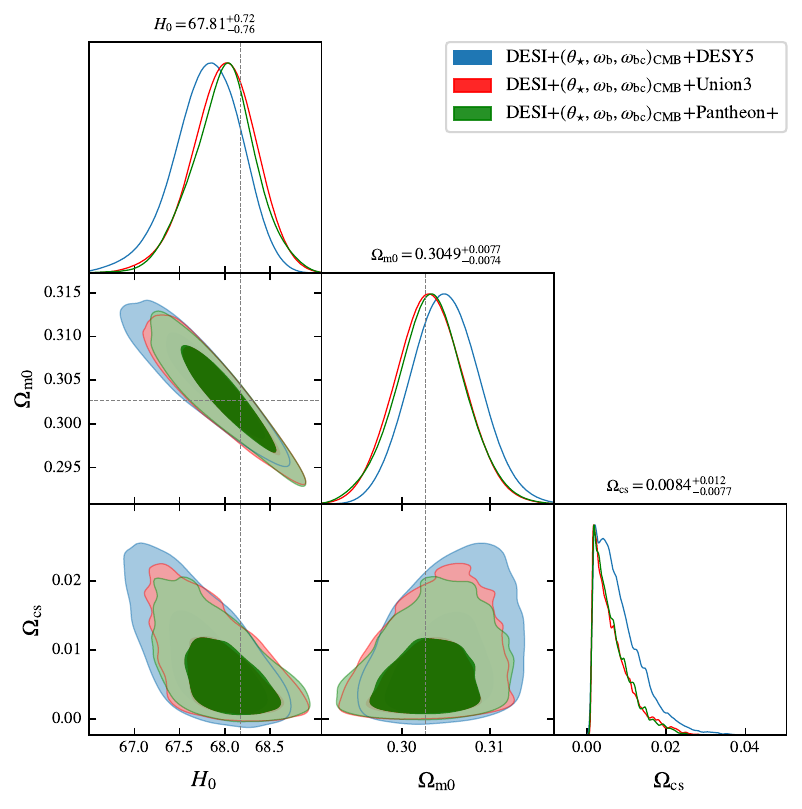}
		\caption{Posterior distributions of parameters in the cosmic string model combined with DESI, CMB and supernovae. The model is with one sampling parameter $\Omega_{\rm cs}$. The blue contours indicate the constraint from DESY5, the red contours correspond to Union3 and the green contours are for the Pantheon+. The gray dashed lines correspond to the value of $H_0$ and $\Omega_{\rm m0}$ of $\Lambda$CDM \cite{DESI:2025zgx}.}\label{cs}
	\end{figure*}

\begin{table}[H]
	\centering
	\renewcommand{\arraystretch}{1.3}
	\begin{tabular}{lccc}
	\toprule
	 & +PantheonPlus & +Union3 & +DESY5 \\
	\midrule
	\multicolumn{4}{l}{\textbf{DESI DR2 BAO + $\left(\theta_{\star},\omega_{\rm b},\omega_{\rm bc}\right)_{\rm CMB}$}} \\
	\midrule
	\multicolumn{4}{l}{\textbf{Domain wall}+$\mathbf{\Lambda}$\textbf{CDM}} \\
	$\Delta \chi_{\rm MAP}^2$ 
		& $-0.064$ 
		& $-0.014$ 
		& $-1.72$ \\
	$ -2\Delta\log \mathcal{P}$ 
		& $-6.05$ 
		& $-6.01$ 
		& $-7.71$ \\
	$\Delta \text{DIC}$ 
		& $0.62$ 
		& $0.73$ 
		& $-0.94$ \\
	$\ln B$ 
		& $-0.69\pm0.34$ 
		& $-0.31\pm 0.33$ 
		& $0.32\pm 0.32$ \\
	\midrule
	\multicolumn{4}{l}{\textbf{Cosmic string}+$\mathbf{\Lambda}$\textbf{CDM}} \\
	$\Delta \chi_{\rm MAP}^2$ 
		& $0.19$ 
		& $0.21$ 
		& $0.037$ \\
	$ -2\Delta\log \mathcal{P}$ 
		& $-5.82$ 
		& $-5.80$ 
		& $-5.97$ \\
	$\Delta \text{DIC}$ 
		& $1.71$ 
		& $1.83$ 
		& $1.40$ \\
	$\ln B$ 
		& $ -2.45 \pm 0.35$ 
		& $ -3.09\pm 0.36$ 
		& $ -2.93\pm 0.35 $ \\
	\midrule
	\multicolumn{4}{l}{$\Omega_k$+$\mathbf{\Lambda}$\textbf{CDM}} \\
	$\Delta \chi_{\rm MAP}^2$ 
		& $-4.55$ 
		& $-4.49$ 
		& $-5.10$ \\
	$ -2\Delta\log \mathcal{P}$ 
		& $-10.54$ 
		& $-10.49$ 
		& $-11.09$ \\
	$\Delta \text{DIC}$ 
		& $-2.88$ 
		& $-2.77$ 
		& $-3.47$ \\
	$\ln B$ 
		& $ 0.10\pm 0.36$ 
		& $ -0.13\pm 0.36 $ 
		& $ -0.25\pm 0.35 $ \\
	\bottomrule
	\end{tabular}
    \label{table1}
	\caption{Fiting result for different models comparing with $\Lambda$CDM.}
	\end{table}
We also present the triangle plot of the posterior distributions of the parameters for different models (see Fig.~\ref{dw} and Fig.~\ref{cs}). The data are analyzed and the plots are generated using \textbf{GetDist}~\cite{Lewis:2019xzd}.

  To identify the best‐fit parameters and their associated $\chi^2$, we perform a minimization with \textbf{iminuit}~\cite{iminuit} initialized at the maximum‐a‐posteriori (MAP) point of each MCMC chain. We calculate $\Delta \chi_{\rm MAP}^2$, which is the reduction in the best-fit $\chi^2$ relative to the $\Lambda$CDM.
\begin{align}
\Delta\chi^2_{\rm MAP}
= -2\,\Delta\ln\mathcal{L}
= -2\bigl[\ln\mathcal{L}_{\rm MAP} - \ln\mathcal{L}_{\rm MAP}^{\Lambda\text{CDM}}\bigr],
\end{align}
Here, $\Delta \chi_{\rm MAP}^2$ is twice the difference in negative log‐likelihood values evaluated at the respective MAP points, and a more negative $\Delta \chi_{\rm MAP}^2$ indicates a better fit. Moreover, we also calculated the difference in the negative logarithms of the posterior probabilities between the corresponding model and the standard model, denoted as $-2\,\Delta\log\mathcal{P}$.

\vspace{2em}

In addition, we examine the deviance information criterion (DIC)~\cite{Spiegelhalter:2002yvw,liddle2007information,DESI:2025fii}, which is defined as
\begin{align}
    \mathrm{DIC}\equiv D(\bar{\theta})+2p_D,
\end{align}
where $\theta$ is the set of sampling parameters, $p_D\equiv \overline{D(\theta)}-D(\bar{\theta})$
The DIC, comparing with \(\chi^2_{\rm MAP}\), accounts for model complexity; models with smaller DIC values are preferred. DIC penalizes models through two components: the posterior mean deviance, \(\bar{D}\), which measures goodness of fit; and the effective number of parameters, \(p_{D}\). Because \(\bar{D}\) typically decreases when additional parameters are included, the \(p_{D}\) term compensates by penalizing complexity, thereby favoring models with fewer effective parameters. 

To compare the models, we compute the Bayes factors using the nested sampler \textbf{PolyChord}~\cite{Handley:2015fda,Handley:2015vkr}, which is integrated into \textbf{Cobaya}. The Bayesian evidence is then estimated using the \textbf{anesthetic} package~\cite{Handley:2019mfs}. The Bayes factor of two models is defined as the ratio of their evidence $\mathcal{Z}$ under a given dataset: $B=\mathcal{Z}_1/\mathcal{Z}_2$, where the evidence (marginal likelihood) is evaluated as
\begin{align}
\mathcal{Z} = \int \mathcal{L}(\theta)\,\pi(\theta)\,\mathrm{d}\theta,
\end{align}
where $\mathcal{L}(\theta)$ is the likelihood and $\pi(\theta)$ is the prior.
And we calculate the logarithm of the evidence, where $\mathrm{ln}B=\Delta\mathrm{ln}\mathcal{Z}$. 
for each model. Unlike the traditional \(\Delta\chi^2\) criterion, which compares only the maximum‐likelihood values and typically assumes nested models and Gaussian errors, \(\Delta\ln \mathcal{Z}\) automatically incorporates the full parameter space volume, thus penalizing unnecessary complexity and implementing Occam’s razor in a principled way. As a result, \(\Delta\ln \mathcal{Z}\) remains valid even for non‐nested models or non‐Gaussian likelihoods, yielding a more robust and probabilistically consistent measure of model preference.

In summary, negative values of \(\Delta\chi^{2}_{\rm MAP}\) indicate that the model achieves a better best‐fit \(\chi^{2}\) than \(\Lambda\)CDM; a negative \(\Delta\mathrm{DIC}\) indicates that the model is preferred according to the Deviance Information Criterion; and \(\ln B > 0\) constitutes Bayes evidence in favor of the model, whereas \(\ln B < 0\) indicates evidence against it, according to the Jeffreys scale. Based on the values of these quantities in Tab.\;\ref{table1}, the corresponding results are discussed as follows:


For the domain-wall model \(\bigl(w=-2/3\bigr)\), we find that, when combined with the Pantheon+ and Union3 supernova samples, it yields a slightly better fit than the $\Lambda$CDM model. When combined with the DESY5 dataset, even a domain-wall contribution at the few-percent level can improve the fit, giving $\Delta \chi^2=-1.72$ compared to $\Lambda$CDM, suggesting a mild preference for the domain-wall model.
Considering the Deviance Information Criterion (DIC), the combination with Pantheon+ and Union3 data yields \(\Delta\mathrm{DIC}\approx+0.6\) to \(+0.7\), indicating that the data does not particularly favor the increased complexity of the model. In contrast, the combination with DESY5 gives \(\Delta\mathrm{DIC}=-0.94\), implying that, after penalizing model complexity, the domain-wall extension is actually favored. A similar trend is also observed in the logarithmic Bayes factor, $\ln B$.

This preference arises because a network of domain walls with equation of state \(w = -2/3\) and present-day fractional density \(\Omega_{\rm dw}\simeq 5\%\) (see Fig.~\ref{dw}) behaves like an additional fluid that redshifts more slowly than matter but more rapidly than a cosmological constant. This enables a transient phase of cosmic acceleration, which can modestly improve the fit to supernova data.
Overall, we find that a model including a few-percent domain-wall component can achieve up to \(\Delta\chi^{2}_{\rm MAP}\simeq -1.72\) when combining BAO, CMB, and supernova datasets (DESY5), suggesting that such a modification to \(\Lambda\)CDM is mildly preferred by current data and merits consideration as a viable extension.

For the cosmic string model, we find \(\Delta\chi^{2}_{\rm MAP} = +0.19\), \(+0.2\), and +0.037 when combined with the Pantheon+, Union3, and DESY5 supernova datasets, respectively. These results indicate that the data does not exhibit a significant preference for the presence of cosmic strings. This conclusion is also supported by Fig.~\ref{cs}, which shows a preference for a low or negligible cosmic string contribution to the energy budget of the universe.
The situation worsens when model complexity is taken into account. The Deviance Information Criterion yields \(\Delta\mathrm{DIC}\approx 1.7, 1.8\), and 1.4 for the Pantheon+, Union3, and DESY5 datasets, respectively, suggesting that the added complexity of the cosmic string model is not justified by the data. A similar conclusion is drawn from the logarithmic Bayes factor $\ln B$. 

A network of long cosmic strings behaves like a fluid with an equation of state \(w=-1/3\), corresponding to an energy density that redshifts as \(\rho_{\rm cs}\propto a^{-2}\). Such a component would contribute a term to the Hubble parameter of the form \(\Omega_{\rm cs}(1+z)^{2}\). Although one might expect this additional contribution to improve the fit—particularly as suggested by Fig.~\ref{fig:hubble_td}—the data does not show a meaningful preference for the existence of cosmic strings in our universe. Note that our result is consistent with that of the recent work \cite{Cheng:2025hug} which studies several case of cosmic strings.

For comparison, we also present results for an alternative extension of the $\Lambda$CDM model: the $\Omega_k + \Lambda$CDM framework. Although both cosmic strings and spatial curvature $\Omega_k$ scale as $a^{-2}$, they are fundamentally different in nature. Spatial curvature is a geometric property that characterizes the overall flatness of spacetime and enters the Friedmann equation as a term evolving like $a^{-2}$. However, its influence goes beyond modifying the expansion rate: while cosmic strings affect only the Hubble parameter $H(z)$, spatial curvature also alters the functional form of distance-related observables, such as the transverse comoving distance.

This distinction is clearly reflected in our fitting results. Compared to the cosmic string model, the $\Omega_k + \Lambda$CDM extension consistently yields $\Delta\chi^2_{\rm MAP} \approx -4$ to $-5$. It also results in $\Delta \mathrm{DIC} \approx -3$, suggesting that, after accounting for model complexity, the curved-universe extension is statistically favored. However, the Bayes factor remains inconclusive, with $\ln B \approx 0 \pm 0.36$, providing neither strong support for nor against the inclusion of spatial curvature. In Bayesian terms, the data are compatible with a small, nonzero $\Omega_k$, but do not compellingly require its presence.

\section{Conclusion}

In this work, we studied the potential contribution of topological defects—specifically domain walls and cosmic strings—to the energy content of the universe as a means to explain the mild tension observed in the DESI DR2 BAO data. Instead of invoking a time-varying dark energy component, we proposed that these defects could mimic dynamical effects through their distinct redshift behavior.

Our analysis, based on a combination of DESI DR2, CMB compressed likelihoods, and various Type Ia supernova datasets, revealed no significant evidence for cosmic strings. In contrast, we found that including a domain wall component at the percent level modestly improves the fit to the data. When combined with the DESY5 supernova dataset, the domain wall model yields a best-fit improvement of $\Delta\chi^2 = -1.72$ over $\Lambda$CDM and shows a slight preference in terms of the DIC and Bayes factor.

This result suggests that domain walls, which redshift more slowly than matter but faster than a cosmological constant, can effectively mimic an evolving dark energy component in the relevant redshift range. While not conclusive, the improvement indicates that topological defects—particularly domain walls—remain a viable and interesting extension to $\Lambda$CDM, meriting further investigation with future cosmological data.

\vspace*{3mm}
\noindent 
\textbf{Acknowledgments}\\[0.5mm] 
The work of H.\,A. is supported in part by
the National Key R\&D Program of China under Grant Nos. 2023YFA1607104 and 2021YFC2203100, and the National Science Foundation of China under Grant No. 12475107. The work of C.\,H.\ is supported by the National Key R{\&}D Program of China under grant 2023YFA1606100 and by the National Natural Science Foundation 
of China under grants No. 12435005. 
C.\,H.\ acknowledges supports from the Sun Yat-Sen University Science Foundation, 
the Fundamental Research Funds for the Central Universities at Sun Yat-sen University under Grant No.\,24qnpy117, and the Key Laboratory of Particle Astrophysics and Cosmology (MOE) of Shanghai Jiao Tong University.\

\bibliographystyle{utphys}
\bibliography{references}

\end{document}